\documentstyle[11pt,aaspp4]{article}

\begin{document}     

\baselineskip 0.216in

\def\today{\ifcase\month\or
January\or February\or March\or April\or May\or June\or
July\or August\or September\or October\or November\or December\fi
\space\number\day, \number\year}


\newcommand{\squig}{$\sim$}
\newcommand{\squigleq}{\mbox{$^{<}\mskip-10.5mu_\sim$}}
\newcommand{\squiggeq}{\mbox{$^{>}\mskip-10.5mu_\sim$}}
\newcommand{\squiggeqmm}{\mbox{$^{>}\mskip-10.5mu_\sim$}}
\newcommand{\decsec}[2]{$#1\mbox{$''\mskip-7.6mu.\,$}#2$}
\newcommand{\decsecmm}[2]{#1\mbox{$''\mskip-7.6mu.\,$}#2}
\newcommand{\decdeg}[2]{$#1\mbox{$^\circ\mskip-6.6mu.\,$}#2$}
\newcommand{\decdegmm}[2]{#1\mbox{$^\circ\mskip-6.6mu.\,$}#2}
\newcommand{\decsectim}[2]{$#1\mbox{$^{\rm s}\mskip-6.3mu.\,$}#2$}
\newcommand{\decmin}[2]{$#1\mbox{$'\mskip-5.6mu.$}#2$}
\newcommand{\asecbyasec}[2]{#1$''\times$#2$''$}
\newcommand{\aminbyamin}[2]{#1$'\times$#2$'$}

\title{Ultracompact X-ray Binaries in Globular Clusters: Variability of
the Optical Counterpart of X1832--330 in NGC\,6652\,\footnote{\ Based
on observations with the NASA/ESA
Hubble Space Telescope, obtained at the Space Telescope Science
Institute, which is operated by the Association of Universities for
Research in Astronomy, Inc., under NASA contract NAS5-26555.}
}
\author{Eric W. Deutsch, Bruce Margon, and Scott F. Anderson}
\affil{Department of Astronomy, 
       University of Washington, Box 351580,
       Seattle, WA 98195-1580\\
       deutsch@astro.washington.edu; margon@astro.washington.edu;
       anderson@astro.washington.edu}


\begin{center}
Accepted for publication in The Astrophysical Journal Letters\\
{\it received 1999 November 5; accepted 1999 December 13}
\end{center}

\begin{abstract}

Evidence is emerging that the luminous X-ray sources in the cores of
globular clusters may often consist of, or perhaps even as a class
be dominated by, ultracompact (P\,\squigleq\,1~hr) binary stars.  To the two
such systems already known, in NGC\,6624 and NGC\,6712, we now add
evidence for two more. We detect large amplitude variability in the
candidate optical counterpart for the X-ray source in the core of
NGC\,6652. Although the available observations are relatively brief, the
existing {\it Hubble Space Telescope} data indicate a strong
43.6~min periodic modulation of the visible flux of semi-amplitude 30\%.
Further, although the orbital period of the source in NGC\,1851 is not
yet explicitly measured, we demonstrate that previous correlations
of optical luminosity with X-ray luminosity and accretion disk size,
strengthened by recent data, strongly imply that the period of that system
is also less than 1~hr.  Thus currently there is evidence that 4 of the 7
globular cluster X-ray sources with constrained periods are ultracompact,
a fraction far greater than that found in X-ray binaries the field.

\end{abstract}

\keywords{globular clusters: individual (NGC 6652) --- stars: neutron ---
ultraviolet: stars --- X-rays: bursts --- X-rays: stars}

\section{INTRODUCTION}

It has long been recognized that highly luminous
($L_X\sim10^{36-37}$~erg~s$^{-1}$) X-ray sources in the cores of globular
clusters are grossly overrepresented with respect to the general galactic
population (Katz 1975; Clark 1975): clusters contain $10^{-4}$ of the
Galaxy's mass, but $10^{-1}$ of the low-mass X-ray binary (LMXB) sources.
A variety of lines of evidence tell us that these objects are neutron
stars in very close binary systems, but the precise mechanisms which
enhance their formation in clusters, and protect them from disruption
thereafter, are obscure.  Essentially all of the bright cluster
sources are also X-ray bursters, frequently emitting $L_X\sim10^4~L_\odot$
in just a few seconds.  Close binaries also have a profound
effect on cluster dynamics: just a few such objects in a cluster have
a store of orbital kinetic energy which can equal or exceed the orbital
energy of all $10^5$ single cluster stars.  Thus the study of these rare
and odd objects also has important macroscopic implications for
the dynamical evolution of clusters (Elson et al. 1987, Hut et al.~1992).

Considerable recent progress in the understanding of the intense
bursting X-ray sources in globular cluster cores is in large part
due to {\it Hubble Space Telescope} ({\it HST}) identifications and
follow-on studies of optical/UV counterparts, and to the realization
that at least two of the cluster sources are exotic, ultra-short period
double-degenerate binary systems: $P=11$ min for X1820--303 in NGC\,6624
(Stella et al. 1987; Anderson et al. 1997) and $P=21$ min for X1850--087
in NGC\,6712 (Homer et al.~1996).
The optical/UV
studies with {\it HST} have in one sense proven highly successful:
a plausible optical counterpart has been identified and/or studied in
detail in each of the clusters carefully scrutinized thus far.  However,
the diversity in properties of the six counterparts now identified is
enormous, with optical luminosities ranging from $M_B$=6 to $M_B$=1,
and confirmed binary orbital periods ranging from 11~min to 17~hr.

The only optical counterpart candidate thus far for X1832--330 in
NGC\,6652 was advanced by Deutsch et al. (1998b; hereafter Paper I).
The object, denoted Star 49, exhibits a UV excess in the {\it HST}
data similar to other known LMXB optical counterparts, and similar
absolute magnitude to the optical counterpart of the LMXB in NGC\,1851.
However, the region surveyed in Paper I does not completely cover the {\it
ROSAT} X-ray error circle derived in that work, and the images are not
very deep.  Furthermore, the position of Star 49 is discrepant
at the $2.3\sigma$ level with the X-ray coordinates.  For these reasons,
Paper I suggests that while Star~49 is the best candidate for the optical
counterpart, its identification remains tentative.

\section{ANALYSIS}


Since the initial search for the optical counterpart and discovery of
Star 49 in Paper I, additional WFPC2 observations have become available
in the {\it Hubble Data Archive}.  Here we discuss three orbits of F555W
(V) and F814W (I) imaging data obtained on 1997 September 5, as well as
one orbit of F555W, F439W (B), and F218W imaging data obtained on 1995
September 13.  In the former observation, seventeen $\sim$20~s F555W
and F814W exposures were taken on the first orbit, twelve 160~s F555W
exposures on the second orbit, and twelve 160~s F814W exposures on the
third.  For the latter observation, fine lock was not achieved and the
stellar images are elongated, the cluster is miscentered on the CCDs, and
the F218W exposures failed completely.  The retake data for this failed
observation were successful, and are discussed in Paper I, but did not
include the proposed optical counterpart.  The early, poor quality data,
however, do actually include Star 49, are usable, and will be briefly
discussed here as they were overlooked in Paper~I.  All these data were
acquired during unrelated programs to study the cluster NGC\,6652 itself.

Despite the suboptimal sampling of the WFPC2 {\it Wide Field} CCDs, on
which the optical counterpart falls in these observations, the cluster
is sufficiently sparse and Star 49 sufficiently unblended that aperture
photometry is entirely adequate to measure magnitudes for this object
and a set of nearby comparison stars.  Aperture corrections are taken
from Table 2(a) in Holtzman et al. (1995b).  The photometric measurements
have not been corrected for geometric distortions, nor is any correction
for charge transfer efficiency losses (Holtzman et al. 1995b) applied;
for most of the images, these effects should contribute errors of only a
few percent.  We use the photometric zero points for the STMAG system from
Table 9 ($Z_{STMAG}$) in Holtzman et al. (1995a).  Systematic errors for
all magnitudes due to uncertainties in detector performance and absolute
calibration are \squig 5\%.

Five nearby reference stars also photometered show no variability to the
limit of the derived uncertainties; three of the five are of comparable
brightness of Star 49.  For Star 49 itself, variability is suggested
in the 1995 epoch observations, and large amplitude variability is
clearly evident in the 1997 epoch observations.  This large-amplitude
variability, when coupled with the UV excess demonstrated in Paper
I, lends considerable confidence that this object is the correct
identification of the optical counterpart to X1832--330.

From the three orbits of F555W and F814W observations, we find
$<m_{555}>=19.48$ and $<m_{814}>=19.90$.  To create a single light
curve of all the data, we subtract 0.4 mag from $m_{814}$ to create
approximately filter-independent magnitudes.  This result is searched
for periodic components using algorithms described in Horne \& Baliunas
(1986).

In Fig.~1a we show the Fourier transform with the CLEAN algorithm
(Roberts et al. 1987) applied to remove the effects of the window
function.  We find the strong peak seen at $43.7\pm0.7$~min to have 99.5\%
significance, based on the original (i.e. prior to CLEANing) periodogram,
using methods in Horne \& Baliunas (1986).  In Fig.~1b we show the entire
light curve ($m_{555}$, $m_{814}-0.4$) for the three orbit observation.
Uncertainty bars are provided for each datapoint, although they are
sometimes smaller than the symbols themselves.  A non-linear least
squares fitting algorithm is used to determine the best fit sinusoid,
which is overplotted on the light curve.  The result is a best-fit
period $43.6 \pm 0.6$~min, semi-amplitude $0.30 \pm 0.05$~mag, and
mean magnitude $19.49 \pm 0.03$.  The sinusoid does describe the broad
trends in the data quite well, but clearly a large amount of aperiodic
flickering is also evident.  In Fig.~1c we show the light curve averaged into
10 phase bins.  Photometric uncertainties are smaller than the symbols.
A few points deviate significantly from the sinusoid fit, most likely
due to the strong flickering and small amount of data.


During 1998 November 28--30, we obtained $\sim35$~ks of X-ray observation
on X1832--330 with {\it Rossi X-ray Timing Explorer}.  The data are
processed through the standard {\it Ftools} package to obtain a calibrated
light curve.  Two Type I X-ray burst events are evident, confirming the
bursting nature of this source first reported by in 't Zand et al. (1998)
with {\it BeppoSAX} observations.  After background subtraction, we
measure a persistent countrate $\sim100$ s$^{-1}$, which is $\sim6$~mCrab,
similar to fluxes reported previously for this object.  A search of the
background-subtracted light curve reveals no significant periodicities,
except for some power at half the {\it RXTE} orbital period, apparently
induced by the calculated background model.  In particular, there does
not appear to be any significant power at the 43.6~min optical period.
The rms scatter in the X-ray light curve is consistent with Poisson noise;
we find no evidence for flickering, which might be expected based on our
optical observations.  However, as the X-ray and optical observations were
made over a year apart, no firm conclusions can be drawn from the lack of
X-ray flickering.  A further analysis of the light curve,
spectra, and bursts in these X-ray data will be discussed elsewhere.

Mukai \& Smale (2000) present an X-ray observation of X1832--330
from a 1996 {\it ASCA} observation.  They also find no periodic X-ray
modulation, but they do provide evidence for X-ray variability with a
similar timescale as the flickering seen in our optical light curve.

\section{DISCUSSION}

The peak in the periodogram of the optical data has 99.5\% significance,
indicating that it is quite improbable that uncorrelated, Gaussian noise
would randomly generate such a strong periodic signal.  However, the
evident scatter in the photometry is not due to measurement uncertainties,
but rather an inherent flickering in the source, and this behavior is
likely to produce significantly correlated ``noise''.  Thus the formal
significance calculation for the periodicity may overestimate the actual
confidence.

Our derived period is close to, but statistically different from, half the
{\it HST} orbital period.  Two further tests increase our confidence that
this behavior is not an artifact of the {\it HST} orbit.  As noted in \S
2.1, several stars near to and of similar magnitude to Star~49 have been
measured from the same data, and show no variability at this or other
periods.  We have also randomized the association of observation times
versus magnitudes for Star~49 and rereduced the data.  Although periods
due to incomplete removal of the window function should then remain, as
the observing times are identical in these randomized data and in the
original observations, the resulting periodograms show no significant
power at 43~min.  We find that only $\sim5\times10^{-4}$
of 10000 trials show a peak at any frequency as high as the 43~min one.

Although the marked variability we report here adds confidence to the
identification of Star~49 with the X-ray source, the mediocre agreement
of the X-ray position with the object still leaves some uncertainty.  This issue
will almost surely be settled by a scheduled observation by the {\it
Chandra X-ray Observatory}, which should yield a highly accurate position.
However, if we accept that our observed optical variations in Star~49 are
indeed periodic, there are few alternatives to identifying this object as
a LMXB, given its measured characteristics, irrespective of the issues
of the X-ray position.  The most extreme SX~Phe stars, for example,
have periods less than 1~hr, but do not display the marked flickering
we observe, so stellar pulsation seems implausible.  If the period is
instead orbital, the flickering implies a mass-transfer system.  However,
no classical cataclysmic variables (CVs) are known with periods less than
1~hr, and although CVs share the colors and flickering of Star~49, they
are typically 3--4~mag less luminous than our object in any case. The
He-rich AM~CVn stars have the appropriate colors, flickering and period
range, but are thought to have $M_V\sim10$ (Warner 1995), so would be
$\sim10^2$ fainter than Star~49.  Thus, given the measured period,
luminosity, colors, and flickering of this object, one would likely
conclude it is an LMXB even without knowledge of the fact that there is
indeed a bright X-ray source observed in the region.

\bigskip

We summarize a variety of parameters for all of the globular cluster LMXB
sources and their host clusters in Table 1.  Cluster data are primarily
compiled from Djorgovski (1993) and other references in the same volume.
Column 8 lists ${\rm F_X}$ values, which we derive by taking a
mean of the {\it RXTE} ASM flux measurements since 1996, and applying a
simple correction to convert approximately to $\mu$Jy.  In column 9, we
apply the distance correction and give an approximate X-ray luminosity
in $10^{36}$ erg s$^{-1}$ for the 2--10 keV ASM band.  The absolute
calibration is only an estimate and should be treated with caution, but
the relative values are likely reliable.  Finally, in column 10 we list
$\xi=B_0+2.5{\rm\,log\,F_X(\mu Jy})$, the parameter used by van Paradijs
\& McClintock (1995) to characterize the ratio of X-ray to optical flux.

That the optical luminosity should depend upon the X-ray luminosity
and the size of the accretion disk has been quantified by van Paradijs
\& McClintock (1994; hereafter vPM94).  They define the parameter
$\Sigma=(L_X/L_{\rm Edd})^{1/2} (P/1\,{\rm hr})^{2/3}$ and find a strong
correlation, such that M$_V=1.57(\pm0.24)-2.27(\pm0.32)\log \Sigma$.
In Fig.~\ref{periodrel} we show a similar figure as in vPM94, but we use
M$_B$ instead of M$_V$, which is likely to be reasonable as vPM94 find an
average $(B-V)_0=-0.09\pm0.14$ for field LMXBs.  The data for globular
cluster LMXBs are derived here and from Deutsch et al. (1998a), and
are plotted with large diamonds, approximately indicating the entire
known range of optical and X-ray luminosity.  The solid line indicates
the best fit to all LMXBs by vPM94.  The dotted lines denote the
apparent full range of possible values (using vPM94's best fit slope).
For NGC\,1851, no orbital period is known, but the optical and X-ray
luminosities are measured.  We therefore draw a dashed line which is likely
to encompass the probable range of orbital periods, $0.2-0.85$ hr, where
the lower bound is taken to be the shortest orbital period known and
the upper value is the maximum period implied by our dotted line bounds.
We thus predict that the orbital period of X0512--401 will prove to be
less than 1 hr.  Based on model accretion disks, Deutsch et al. (2000)
also infer that X0512--401 must have orbital period less than one hour.

As listed in Table 1, an eclipse period of 12.4~hr was recently reported
by in 't Zand et al. (2000) for the LMXB in the globular cluster
Terzan 6.  Using behavior exhibited by the GC LMXB sources in Fig.~2,
we can now infer that the optical counterpart of that source (for
which there has not yet been a search) will have $M_B\sim2\pm1$.
The high inclination ($i>74^\circ$) inferred from the eclipse behavior
by in 't Zand et al. (2000) suggests that the luminosity may well be
at the fainter end of the above range, and thus similar to the optical
counterpart in NGC\,6441.  However, the high reddening to Terzan~6 will
making discovery of the optical counterpart by conventional means ({\it
HST} observations of UV-excess) extremely difficult (Deutsch et al. 1998b).
A search for eclipses with infrared imaging of the X-ray error circle
may be the easiest method of isolating the optical/IR counterpart of
this source.

\clearpage
\section{CONCLUSIONS}

The ultraviolet-excess candidate for the optical counterpart of the
intense X-ray source in NGC\,6652 suggested by Deutsch et al. (1998b)
is found to be highly photometrically variable. Although the data are of
limited length, the evidence is strong that a significant component of
the variability is periodic, with $P=43.6$~min, most likely the orbital
period of the system.  Regardless of whether or not the variability
is periodic, the marked amplitude of the variations significantly
strengthens the case for the identification of the object with the
bursting X-ray source.  Somewhat tempering this conclusion is the poor
positional agreement of the object with the only extant X-ray data,
although {\it Chandra X-ray Observatory} observations will almost
surely settle this issue. The lack of similar X-ray variability is
inconclusive. Star 49 is clearly unusual regardless of its association
with the X-ray source, but the probability of two such unrelated objects
falling within a few arcseconds of each other is presumably modest, so
we favor the identification of the star and the X-ray burster, pending
the {\it Chandra} data.

We examine a homogeneous set of {\it HST} data on globular cluster X-ray
source counterparts, including Star 49 in NGC\,6652, and find that they
fit well with the correlation of optical luminosity, X-ray luminosity,
and accretion disk size previously discussed by vPM94.  Even if the 43~min
period is not confirmed, the orbital period of X1832--330 must still be
less than $\sim2$~hr if it is to follow the relation of this diagram.
Using a somewhat different argument, Mukai \& Smale (2000) also infer
that X1832--330 in NGC\,6652 is a short period system.

The correlation in this diagram also strongly implies that the X-ray
source in NGC\,1851 must have orbital period $P<1$~hr as well. A similar
conclusion is reached by Deutsch et al. (2000) via an independent
argument, through examination of the spectral energy distribution of
that object.  Thus four of the seven central globular cluster X-ray
sources where orbital periods are constrained or known are inferred to
be ultracompact, a fraction considerably in excess of that for field low
mass X-ray binaries; only $\sim7$\% of field LMXBs with known periods
have $P<1$~hr in the compilation of van Paradijs (1995).  In fact,
if ultracompact systems in GC LMXBs were as rare as in the field, then
the binomial probability that at least four out of seven systems are
by chance found to be ultracompact is only $7\times10^{-4}$.  One can
readily imagine multiple explanations for this significant overabundance
of compact systems, although the unique dynamic environment of cluster
cores is certainly a most tempting factor to invoke.  It is not clear
whether the same explanation applies even to all members of this small
sample (the double-degenerate systems in NGC\,6624 and NGC\,6712 may be
unique), or that observational selection may be at work.

\acknowledgments

{\it RXTE} ASM data products were provided by the ASM/RXTE teams at MIT and
at the {\it RXTE} SOF and GOF at NASA's GSFC.  Support for this work
was provided by NASA through grants NAG5-7330 and NAG5-7932, as well as
grant AR-07990.01 from the ST\,ScI, which is operated by AURA, Inc.


\begin{deluxetable}{llccrrrrrr}
\tablewidth 6.8in
\tablenum{1}
\tablecolumns{10}
\tablecaption{X-ray source and corresponding globular cluster data}
\tablehead{
\colhead{Source} &
\colhead{Cluster} &
\colhead{E($B-V$)} &
\colhead{$(m-M)_0$} &
\colhead{[Fe/H]} &
\colhead{$P_{\rm orb}$\,\tablenotemark{a}} &
\colhead{$M_B$} &
\colhead{$F_X$\,\tablenotemark{b}} &
\colhead{$L_X$\,\tablenotemark{c}} &
\colhead{$\xi$\,\tablenotemark{d}}
}
\startdata 
X0512$-$401 & NGC 1851 &  0.02 &  15.43 & $ -1.29$ &  $<1$ &  5.60 &    5.5 &    1.9 & 22.88\nl
X1724$-$307 & Terzan 2 &  1.42 &  14.37 & $ -0.25$ &       &       &   33.6 &    4.3 &      \nl
X1730$-$335 & Liller 1 &  3.00 &  14.68 & $  0.20$ &       &       &   12.8 &    2.2 &      \nl
X1732$-$304 & Terzan 1 &  1.64 &  13.85 & $ -0.71$ &       &       &    6.4 &    0.5 &      \nl
X1745$-$203 & NGC 6440 &  1.00 &  14.64 & $ -0.34$ &       &       &    5.5 &    0.9 &      \nl
X1745$-$248 & Terzan 5 &  1.87 &  14.50 & $ -0.28$ &       &       &        &        &      \nl
X1746$-$370 & NGC 6441 &  0.45 &  15.15 & $ -0.53$ &  5.70 &  2.43 &   28.8 &    7.6 & 21.22\nl
X1747$-$313 & Terzan 6 &  2.04 &  14.16 & $ -0.61$ & 12.36 &       &   31.8 &    3.4 &      \nl
X1820$-$303 & NGC 6624 &  0.27 &  14.54 & $ -0.37$ &  0.19 &  2.99 &  269.5 &   40.6 & 23.61\nl
X1832$-$330 & NGC 6652 &  0.10 &  14.85 & $ -0.99$ &  0.73 &  5.59 &   10.9 &    2.2 & 23.03\nl
X1850$-$087 & NGC 6712 &  0.46 &  14.16 & $ -1.01$ &  0.33 &  4.48 &    7.8 &    0.8 & 20.87\nl
X2127+119   & NGC 7078 &  0.09 &  15.11 & $ -2.17$ & 17.10 &  0.66 &   13.9 &    3.5 & 18.63\nl
\enddata

\tablenotetext{a}{\ X-ray source orbital period in hours}
\tablenotetext{b}{\ Average {\it RXTE} ASM X-ray flux ($2-10$ keV)
    since 1996, approximately calibrated to units of $\mu$Jy}
\tablenotetext{c}{\ Average X-ray luminosity from {\it RXTE} ASM, approximately
    calibrated to units of $10^{36}$ erg~s$^{-1}$ ($2-10$ keV)}
\tablenotetext{d}{\ $\xi=B_0+2.5{\rm\,log\,F_X\,(\mu Jy})$}
\end{deluxetable}


\begin{figure}
\plotone{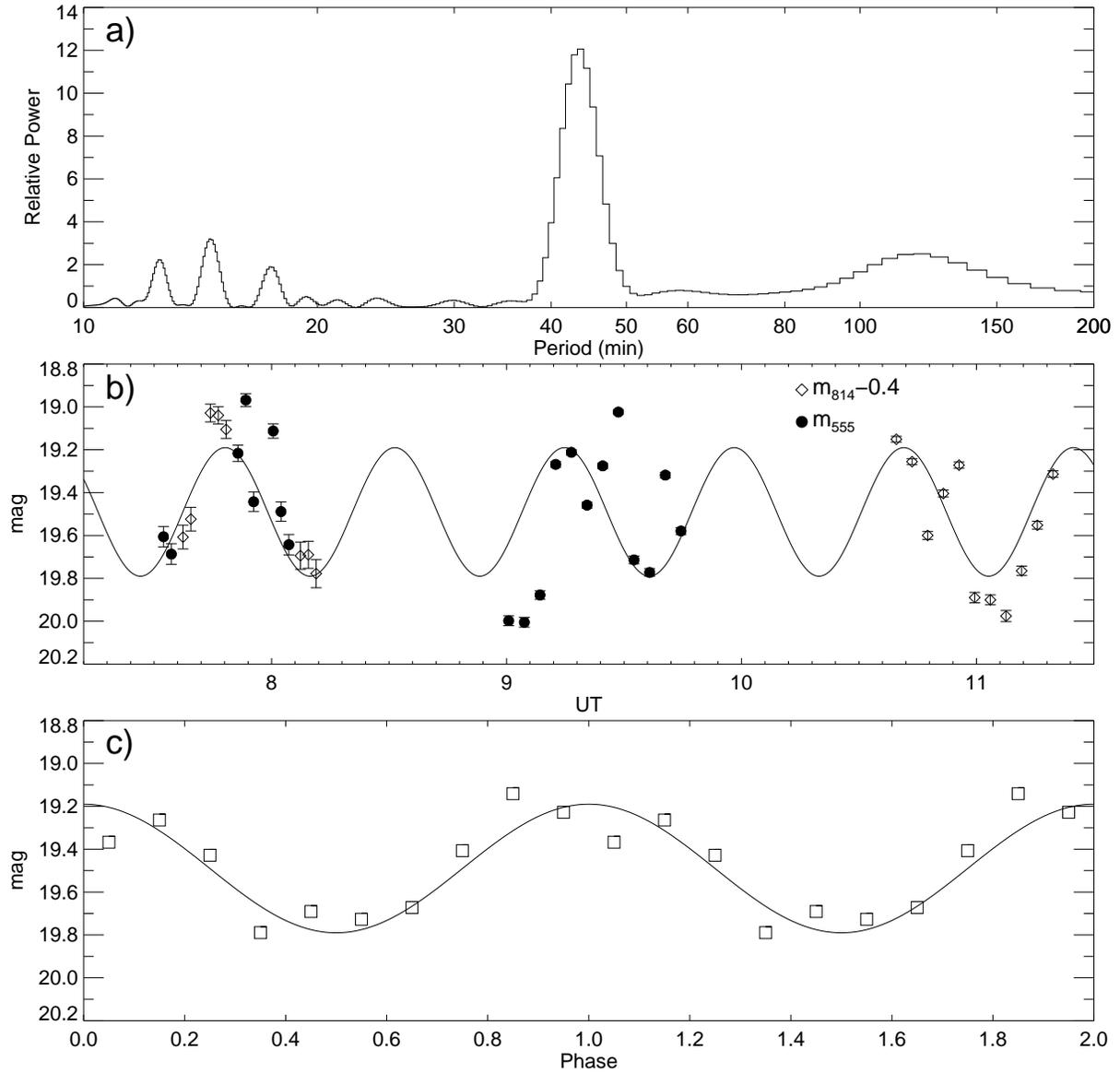}
\caption{Variability analysis of the optical counterpart of the LMXB
in NGC\,6652 based on the three-orbit {\it HST} dataset.  {\it (top)}:
CLEANed Fourier transform of the light curve showing a single significant
peak at 43.6 min.  {\it (center)}: Light curve for the object, with the
best-fitting sinusoid overplotted.  The solid circles denote $m_{555}$
measurements, and the open diamonds, $m_{814}-0.4$, so shifted
to correct for the color of the object.  Photometric uncertainties
($1\sigma$ relative) are shown, but are often smaller than the symbols.
{\it (bottom)}: Light curve folded into 10 phase bins about the 43.6 min
period.  Flickering distorts the true shape of the periodic modulation,
as only a few orbits are sampled.}
\end{figure}

\begin{figure}
\plotone{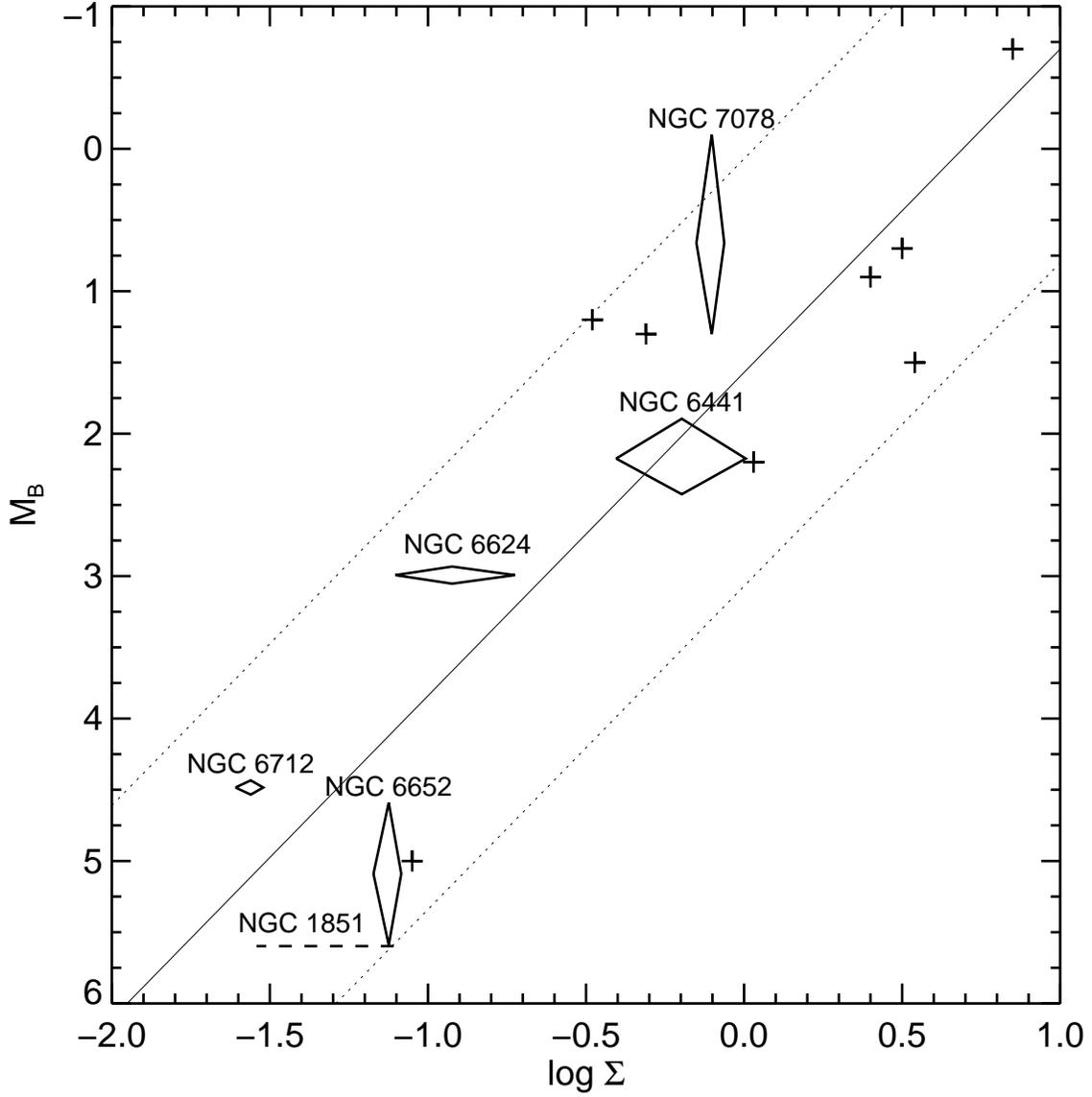}
\caption{Absolute magnitude of LMXBs versus $\Sigma=(L_X/L_{\rm
Edd})^{1/2} (P/1\ {\rm hr})^{2/3}$.  Field LMXBs from van Paradijs \&
McClintock (1994) are denoted with ``+'' symbols.  Globular cluster
LMXBs are displayed with diamonds to indicate known variability in X-ray
or optical luminosity.  The solid diagonal line indicates best fit by
Paradijs \& McClintock; the dotted lines indicate a range which
includes nearly all sources.  Based on these dotted line boundaries,
we predict an orbital period for the source in NGC\,1851 of less than
0.85 hr; the current constraints are denoted with the dashed line.}
\label{periodrel}
\end{figure}

\end{document}